\documentstyle[12pt,psfig]{article}

\begin{document}

\newcommand{\be}{\begin{equation}}
\newcommand{\ee}{\end{equation}}
\newcommand{\bea}{\begin{eqnarray}}
\newcommand{\eea}{\end{eqnarray}}
\newcommand{\vp}{\varphi}
\newcommand{\pd}{{\partial}}
\newcommand{\srr}{1/r^{2}}

\centerline{\Large \bf {Anisotropic Stars II: Stability}}

\vspace{1.0cm}

\centerline{Krsna Dev }
\centerline{{\it Department of Physics and Astronomy,Oberlin College,}}
\centerline{{ \it Oberlin, OH 44074 USA.}}
\vspace{0.5cm }
\centerline{ Marcelo Gleiser}
\centerline{ \it {Department of Physics and Astronomy,Dartmouth College,}}
\centerline{{ \it Hanover, NH 03755 USA.}}

\vspace{1.0cm}
  
\centerline{Abstract}
\noindent We investigate the stability of self-gravitating 
spherically symmetric anisotropic
spheres under radial perturbations. We
consider both the Newtonian and the full 
general-relativistic perturbation
treatment. In the general-relativistic case, we extend the variational
formalism for spheres with isotropic pressure developed by
Chandrasekhar. We find
that, in general, when the tangential pressure is greater than the radial
pressure, the stability of the anisotropic sphere is enhanced when compared
to isotropic configurations. 
In particular, anisotropic spheres are found to be stable for smaller values of the
adiabatic index $\gamma$.

\noindent {\bf KEYWORDS: Radial Perturbations, Stars, Anisotropic Pressure}

\section{Introduction}
\noindent In a recent paper \cite{KRSNA},
we presented a series of new exact solutions of
the Einstein field equations
for self-gravitating, general-relativistic spheres 
with anisotropic pressure. We have found that the presence of
pressure anisotropy, for a large
variety of {\it ansatze} for its functional form, has
important implications for the physical properties of self-gravitating
objects. Namely, both the object's critical mass and surface redshift are modified,
and may violate well-known bounds for isotropic objects ($2M/R < 8/9$ and $z_s \leq 2$).
We have shown that this is true not only for stars of constant energy density,
but also for objects with $\rho\propto 1/r^2$, often used to model
neutron star interiors. Given the fact that pressure
isotropy is an assumption not required by imposing spherical symmetry,
it is clearly of great relevance to investigate if, indeed, these
anisotropic configurations are stable against radial perturbation and, thus,
better candidates to exist in Nature.

The aim of the present paper 
is, then, to develop a formalism which can be used to test the
stability of anisotropic spheres against small radial perturbations.
Our formalism is a generalization of the variational
principle used for investigating the stability properties of isotropic objects.
We reduce the stability analysis to an eigenvalue problem, where the eigenvalues are the
frequencies of the radial modes. 

The dynamical stability of isotropic spheres has been extensively studied
by various authors \cite{CHANDRA, FOWLER, COCKE, THORNE,GLEISER}. A calculation
based on the concept of extremal energy was presented by Fowler \cite{FOWLER}.
Cocke \cite{COCKE}  performed a calculation based on the method 
of extremal entropy. Gleiser, and Gleiser and Watkins applied Chandrasekhar's
variational method to investigate the stability of boson stars, self-gravitating
spheres of complex scalar fields, which are naturally anisotropic \cite{GLEISER}.

Chandrasekhar considered the dynamical stability of isotropic spheres 
as an eigenvalue problem \cite{CHANDRA}.
 He used an analytical approach to compute the 
eigenfrequencies of radial oscillations for isotropic spherical stars. 
The study of the  stability of a star thus becomes a Sturm-Liouville problem.

The main result
of these studies is that, for dynamical stability in general relativity,
isotropic spheres must have an adiabatic index (or exponent)
\be
\gamma  \ge  \frac{4}{3}  + \kappa \frac{M}{R},
\ee

\noindent where $\kappa$ is a number of order unity, that depends on the 
structure of the star, and $M$ and $R$ are the star's
mass and radius, respectively. For white dwarfs, $ \kappa = 2.25$.

The stability of anisotropic spheres in general relativity was studied
numerically by Hillebrandt and Steinmetz \cite{HILL}. An analytical approach in
the spirit of Chandrasekhar's work for isotropic spheres, 
however, does not seem to exist for anisotropic spheres. Our goal is to
obtain this approach.

This paper is organized as follows. In the next section, we obtain
exact solutions for several examples of
Newtonian anisotropic spheres, and study their stability
properties. This will give us some insight into the effects of anisotropy
on the stability of self-gravitating objects. We then proceed,
in section 3, to derive the full general-relativistic
perturbation formalism for anisotropic spheres. In
section 4, we apply the formalism to anisotropic spheres of constant energy density.
In section 5, we apply it to anisotropic spheres with $\rho\propto 1/r^2$. In
both sections, we
follow the exact solutions derived in \cite{KRSNA}. In section 6 we
summarize our results, and discuss possible avenues for future work.

\section{Newtonian  Anisotropic Spheres }
\subsection{Exact solutions for Newtonian Anisotropic Spheres}
We consider  the dynamics of anisotropic spheres under the
influence of Newtonian gravity. The equation of hydrostatic equilibrium 
 with anisotropic pressure in Newtonian gravity is
\be
\label{New}
p_{r}^{\prime}  = - \frac{m(r)\rho(r)}{r^{2}} + \frac{2}{r}
\left(p_{t} - p_{r} \right)
\ee
\noindent where $p_{r}$ is the radial pressure, $p_{t}$
 is the tangential pressure, $\rho$ is the energy density and 
\be
\label{mass22}
m(r) = 4 \pi \int_{0}^{r} \rho(r^{\prime}) r^{\prime 2} dr^{\prime},
\ee

\noindent is the mass contained in a sphere of radius $r$.
This equation may be obtained as the Newtonian limit of the generalized 
Tolman-Oppenheimer-Volkov equation for general relativistic hydrodynamical
equilibrium, or it may be derived using the principles of Newtonian fluid
 mechanics.

\noindent The pressure in isotropic  spheres with constant density,  
$\rho_{o}$,
is given by
\be
p_{r}  =  \frac{2 \pi}{3} {\rho_{0}}^{2}(R^{2} - r^{2}).
\ee
\noindent We note that, in Newtonian gravity, the pressure at the center 
of a sphere with constant density and isotropic pressure can only become 
infinite if the radius of the sphere is infinite.

We will now solve eq. \ref{New}  for various ansatze connecting $p_{r}$ and
$p_{t}$ at constant density $\rho_{0}$. These ansatze are chosen so as 
to correspond to the choices we
will make for the full general relativistic cases. \\

\noindent ${\bf Case ~~I}$: $p_{t} - p_{r} = C  \,{\rho}_{0}^{2} \,r^{2} $  \\
This ansatz assumes that the anisotropy term in eq. (\ref{New})
is proportional to the first term on the right hand side of the equation, 
${\it i.e.}$,
the anisotropy is chosen to mimic the behavior of the purely gravitational
term.   
This ansatz can be interpreted as the Newtonian limit of the 
ansatz that Bowers and Liang used to solve the full general relativistic TOV 
equation \cite{BOWERS}. With this ansatz  eq. (\ref{New}) becomes
\be
p_{r}^{\prime}  = -\frac{4 \pi}{3} {\rho}_{0}^{2} r + 2C {\rho}_{0}^{2} r
\ee
\noindent and the  solution  is 
\be
p_{r} =  {\rho}_{0}^{2}\left( \frac{2 \pi}{3} - C \right)
\left( R^{2} - r^{2} \right) 
\ee

\noindent Since we are considering spheres with constant energy density 
$\rho_{0}$, from eq. (\ref{mass22}),
\be
m(r) =  
\frac{4}{3} \pi \rho_{0} r^{3}~.
\ee
\noindent Therefore, here we can also write
\be
p_{r}= {\rho}_{0}\left( \frac{1}{4} - \frac{3C}{8 \pi} \right)
\left( \frac{2M}{R} - \frac{2m}{r} \right)~.
\ee

\begin{figure}
\hspace{1.5in}
	\psfig{figure=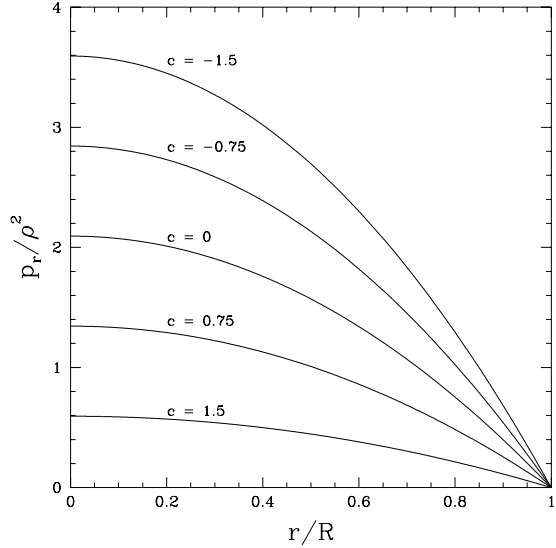,width=3.0in,height=3.0in}
	\caption[$ p_{r}$ vs $r $ for  $p_{t} - p_{r} = C \rho_{0}^{2} r^{2}$]
{\label{figoone} Radial pressure $p_{r}$  as a function of $r$ for the ansatz 
 $p_{t} - p_{r} = C \rho^{2} r^{2}$, parameterized by values of $C$.}
\end{figure}  

\noindent Comparing this solution with the isotropic case ($C=0$), we see that the 
radial pressure has
the same spatial behavior in both cases, and they differ only by a 
multiplicative factor that depends on the amount of anisotropy $C$. If we 
define an effective density:
\be
\bar{\rho} = \left(1 - \frac{3}{2 \pi}C \right)^{1/2} \rho_{0}~,
\ee
\noindent then  we can write
\be
p_{r} =   \frac{2 \pi}{3}{\bar{\rho}}^{2}
\left( R^{2} - r^{2} \right). 
\ee
\noindent Thus, in this model the effect of the anisotropy can be considered as 
a scaling of the density of the sphere. We can take this scaling interpretation
a step further by reintroducing into the expression for $p_{r}$ the 
gravitational constant $G$:
\be
p_{r} =  {\rho}_{0}^{2} G\left( \frac{2 \pi}{3} - C \right)
\left( R^{2} - r^{2} \right).
\ee

\noindent We now define 
\be
\bar{G} = \left( 1 - \frac{3C}{2 \pi}\right)G
\ee

\noindent and find that 
\be
p_{r} =  \frac{2 \pi}{3}{\rho_0}^{2} \bar{G}\left( R^{2} - r^{2} \right).
\ee
\noindent Thus, we can also interpret anisotropy as a variation of the
gravitational constant. This  explains why positive values of $C$ and
hence smaller values of $\bar{G}$ lead to smaller values of the radial 
pressure. For smaller $\bar{G}$ the gravitational force between the particles 
in the sphere is decreased and this leads to a decrease in the radial 
pressure needed to stabilize the sphere. Negative values of $C$ 
have the opposite effect, i.e., $\bar{G}$
is increased and correspondingly $p_{r}$ is also increased.
We plot the radial pressure $p_{r}$ as a function
of the radius $r$ for several values of the anisotropy $C$ in figure 
 \ref{figoone}.

We note that, for  $C = 2 \pi/3$ (${\bar G}=0$), the radial pressure 
vanishes and becomes negative if $C > 2 \pi/3$; in this case, 
no bound solutions are possible. It is interesting to note
that  the solution with $C= 2 \pi/3$ has the following form
\be
p_{r} = 0 ~~~~~~~~~~~~~~ p_{t} = \frac{2 \pi}{3} \rho_{0}^{2} r^{2}.
\ee
\noindent Hence, for this particular solution, the sphere 
is held together by purely tangential stresses. The particles that 
constitute the sphere  are considered to be in random circular orbits 
\cite{HERRERA}. \\

\noindent ${\bf Case ~~II}$: $p_{t} - p_{r} = C \,{\rho}_{0}
 \,p_{r} \,r^{2} $ \\

The solution to eq. (\ref{New}) with this ansatz is 
\be
p_{r} = \frac{2 \pi}{3C} \rho_{0}
 \left[ 1 - e^{-C \rho_{0} \left(R^{2} - r^{2} \right
)}\right].
\ee
For this solution, the pressure at the center is 
\be
p_{c} = \frac{2 \pi}{3C} \rho_{0} \left[ 1 - e^{-C \rho_{0} R^{2}}
\right],
\ee
\noindent and all values of $C$ are allowed. For $C$ small,
\be
p_{r}  =  \frac{2 \pi}{3} {\rho}_{0}^{2}(R^{2} - r^{2}),
\ee
\noindent and
\be
p_{t}  =  \frac{2 \pi}{3} {\rho}_{0}^{2}(R^{2} - r^{2})(1 + C \rho_{0} r^{2}).
\ee
\noindent ${\bf Case ~~III}$: $p_{t} - p_{r} = C \, p_{r}^{2} \,r^{2} $ \\

Here the solution has two distinct forms depending on whether $C$ is greater
than or less than zero. For $C < 0$, the solution is 
\be
p_{r} = \rho_{0} \left(\frac{2 \pi}{3 |C|}\right)^{1/2} \tan
 \left[ \rho_{0} \left(
\frac{2 \pi |C|}{3}\right)^{1/2}
\left(R^{2} - r^{2} \right)  \right],
\ee 
\noindent with
\be
p_{c} = \rho_{0} \left(\frac{2 \pi}{3 |C|}\right)^{1/2}
 \tan \left[ \rho_{0} \left(
\frac{2 \pi |C|}{3}\right)^{1/2}R^{2} \right].
\ee 

\begin{figure}[h !]
\hspace{1.5in}
	\psfig{figure=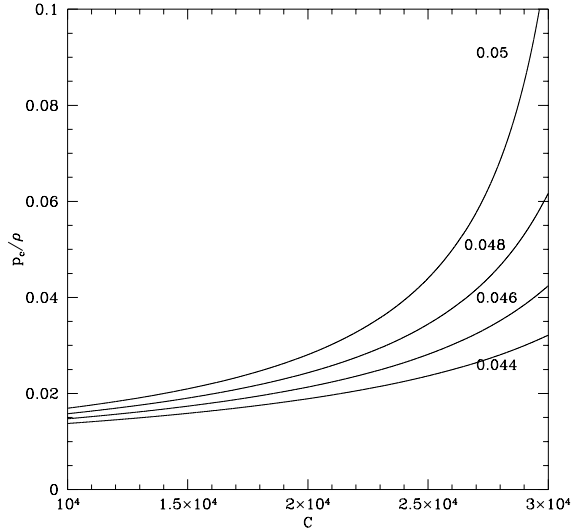,width=3.0in,height=3.0in}
	\caption[$p_{c}$ vs $C$ for  $p_{t} - p_{r} = C {p_{r}}^{2} r^{2}$]
{\label{figtwoo}
 Core  pressure $p_{c}$  as a function of $C$ for the ansatz 
 $p_{t} - p_{r} = C {p_{r}}^{2} r^{2}$, parameterized by values of $2M/R$.}
\end{figure}  

\noindent Thus $p_{c}$ becomes infinite if $ R^2 = \frac{1}{\rho_{0}} 
\left( \frac{ 3 \pi}{8 |C|} \right)^{1/2}$, 
a result that is not possible for Newtonian isotropic spheres with constant
density. However, we note that the values of $C$ for which $p_{c}$ becomes
infinite in Newtonian gravity  are quite large, on the order of $10^{4}$ for $2M/R = 0.05$. We will
see that for the general relativistic case when $2M/R \sim 1 $, the values of
 $C$ for which $p_{c}$ becomes infinite are  of order $1$.
We plot the core pressure $p_{c}$ as a function of $C$ for various 
values of $2M/R$ in figure \ref{figtwoo}.

When $C > 0$ the solution is 
\be
p_{r} = \left( \frac{2 \pi}{3C} \right)^{1/2} \rho_{0} \left[ \frac
{  \exp \left[ \left( \frac{2 \pi C}{3} \right)^{1/2}
\rho_{0} \left(r^{2} - R^{2}
\right) \right]  -1}
{  \exp \left[ \left( \frac{2 \pi C}{3} \right)^{1/2} \rho_{0}
\left(r^{2} - R^{2}
\right) \right]  +1} \right].
\ee
\noindent For this solution $p_{c}$ is always positive and finite.
\\
\subsection{Stability of Newtonian Anisotropic Spheres}

We now proceed to investigate the effects of
small perturbations on the solutions obtained
above. This study is important since it allows us to compute the frequencies
and normal modes of oscillations, enabling us to establish 
 the dynamical stability of our solutions.
We will follow closely the formalism  outlined in 
Shapiro and Teukolsky \cite{SHAPIRO} for isotropic spheres, modifying it where
necessary to the anisotropic case.

It is useful to distinguish between Eulerian and Lagrangian perturbations of 
the fluid variables. If $Q(\vec{x},t)$ is any fluid variable, the Eulerian 
change relative to the unperturbed value $Q_{0}(\vec{x},t)$ is defined as
\be
\delta Q(\vec{x},t)  = Q(\vec{x},t) - Q_{0}(\vec{x},t).
\ee
The Lagrangian change is defined as
\be
\Delta Q(\vec{x},t)  = Q[\vec{x} + \vec{\xi}(\vec{x},t),t] - Q_{0}(\vec{x},t)
\ee
\noindent where $\vec{\xi}(\vec{x},t)$ is an infinitesimal displacement of the 
fluid element. 

The Eulerian approach considers changes in the fluid variables at a particular
point in space, whereas the Lagrangian approach considers changes in a 
particular fluid element. The relationship between the two is
\be
\Delta = \delta + \vec{\xi} \cdot \vec{\nabla}.
\ee

The following equations  govern the dynamics of the unperturbed system:

\begin{enumerate} 

\item The continuity equation that connects the density $\rho$ and velocity $v$,
\be
\frac{\partial \rho}{\partial t}  + ( \rho v)^{\prime} = 0~;
\ee

\item The momentum equation
\be
\frac{dv}{dt} = - \frac{1}{\rho} p_{r}^{\prime} - {\Phi}^{\prime}
 + \frac{2}{r}(p_{t} - p_{r})~, 
\ee 
\noindent where
\be
\frac{d}{dt} = \frac{\partial}{\partial t}  + v \frac{d}{dr}~;
\ee

\item  Poisson's equation, the equation that determines 
 the gravitational potential $\Phi$,
\be
\frac{1}{r^{2}}(r^2 {\Phi}^{\prime})^{\prime} = 4 \pi \rho ~~.
\ee

\end{enumerate}

\noindent We  have adopted  spherical symmetry since we are considering radial 
perturbations.

A Lagrangian perturbation of   the momentum equation gives,

\be
\Delta \left(\frac{dv}{dt}  + \frac{1}{\rho} p_{r}^{\prime} + {\Phi}^{\prime}
 - \frac{2}{r}(p_{t} - p_{r})  \right) = 0. 
\ee 

We note the following:

\begin{itemize}

\item From the continuity equation it follows that 

\be
\Delta \rho  = -\rho \frac{1}{r^{2}} (r^{2} \xi)^{\prime}.
\ee

\item A perturbation of Poisson's equation gives 
 
\be
 (\delta \Phi)^{\prime} = - 4 \pi \rho \xi.
\ee 

\item  The adiabatic exponent $\gamma$  is defined using the following 
expression,

\be
\label{gamma}
\Delta p_{r} \equiv p_{r} \gamma \frac{\Delta \rho}{\rho}.
\ee

\item We also find it convenient to introduce the following symbol,
\be
\Pi \equiv  p_{t} - p_{r}.
\ee

\end{itemize}

\noindent Using eqs. (27)-(30) we can evaluate each term of the  
perturbed momentum equation:

\be
\label{pert1}
\Delta \frac{dv}{dt} = \frac{d^{2} \xi}{d t^{2}} ~~,
\ee

\be
\Delta \left( \frac{1}{\rho} p_{r}^{\prime} \right) = 
-\frac{\Delta \rho}{{\rho}^{2}} p_{r}^{\prime}+ \frac{1}{\rho}
 \Delta p_{r}^{\prime}   = 
\frac{2}{\rho r} \xi p_{r}^{\prime}  + \frac{1}{\rho} \left[-\gamma p_{r}
\frac{1}{r^{2}}(r^{2} \xi)^{\prime} \right]^{\prime} ~~,
\ee

\bea
\Delta ( \Phi)^{\prime} = \frac{d}{dr} + \xi \nabla^{2} \Phi - \frac{2}{r}
\xi \frac{d \Phi}{dr}  
= \frac{2 \xi}{ \rho r} p_{r}^{\prime} - \frac{2}{\rho r^{2}} \Pi \xi  ~~,
\eea

\noindent and

\be
\label{pert2}
\Delta \left( \frac{2}{\rho r} \Pi \right) = 
 \frac{2}{\rho r} \frac{1}{r^{2}} (\xi r^{2})^{\prime} \Pi  
 - \frac{2}{\rho r^{2}} {\xi} \Pi 
 +  \frac{2}{\rho r} \delta \Pi.
\ee
   
\noindent Combining all terms in eqs. (\ref{pert1})- (\ref{pert2}),
 we find that 
radial perturbations are governed by the following equation:

\be
\ddot{\xi} 
- \frac{1}{\rho} \left[\gamma p_{r}
\frac{1}{r^{2}}(r^{2} \xi)^{\prime} \right]^{\prime} 
+\frac{4}{\rho r} \xi p_{r}^{\prime}
-\frac{6}{\rho r^{2}}  \xi \Pi  
-\frac{2}{\rho r}(\xi)^{\prime} \Pi 
- \frac{2}{\rho r} {\xi} \Delta\Pi = 0 ~.
\ee

We now assume that all variables have a time dependence of the form
$ e^{i \omega t}$.
Substituting this form of the time dependence in the above equation, 
 we arrive at an eigenvalue equation
for radial oscillations of a Newtonian spherical star;

\be
\label{evvg}
\left[\gamma p_{r}
\frac{1}{r^{2}}(r^{2} \xi)^{\prime} \right]^{\prime} 
-\frac{4}{ r} \xi p_{r}^{\prime}  + \frac{6}{\rho r^{2}}  \xi \Pi     
+\frac{2}{r}(\xi)^{\prime} \Pi 
+  \frac{2}{ r} {\xi} \Delta \Pi
+ \rho \omega^{2} \xi =0 ~.
\ee

The boundary conditions for this equation are
\be
\label{bc1}
 \xi  = 0 ~~~~~~  {\rm at} ~~~~~ r = 0,
\ee
\be  
\label{bc2}
\Delta p_{r} = 0 ~~~~~~  {\rm at} ~~~~~~ r = R.
\ee
Equation (\ref{evvg}) subject to the boundary conditions (\ref{bc1})  and 
(\ref{bc2})  is a Sturm-Louville eigenvalue problem for $ \omega^{2}$. 
The general theory of these equations gives the following results \cite{MORSE}:
\begin{enumerate}
\item  The eigenvalues are real and form an infinite discrete sequence, \\ 
$~~~~~~~~~~~~~~~~~~{\omega}_{0}^{2} ~ \le ~ {\omega}_{1}^{2} ~
 \le ~ {\omega}_{2}^{2}$ ........... 

\item The $\xi_{n}$ are orthogonal with a weight function $\rho r^{2}$: \\
$ ~~~~~~~~~~~~~~~~~~~~~ \int_{0}^{R} \xi_{n} \xi_{m} \rho r^{2}dr = 0, ~~~
m \neq n.$

\item The $\xi_{n}$ form a complete basis for any function satisfying the 
boundary conditions \ref{bc1} and $\ref{bc2}$.
\end{enumerate}

An important consequence of these results is that, 
if the fundamental mode of the
star is stable  ($ \omega^{2}_{0} \ge 0 $), then all radial modes are stable.
Conversely, if the star is radially unstable, the  fastest growing 
instability will be via the fundamental mode ($\omega_{0}^{2}$ more negative
than all other $\omega^{2}_{n}$).

Equation (\ref{evvg}) can be solved for $\omega^{2}$. 
Multiplying by $ \xi r^{2} $
and integrating from $0$  to $R$ we find
\be
\label{omega}
\omega^{2} = \frac{ \int_{0}^{R} \left\{ \gamma p_{r}
\frac{1}{r^{2}}(r^{2} \xi)^{\prime  2}  +
+ 4 r \xi^{2} p_{r}^{\prime} -6 {\xi}^{2} \Pi       
- 2 r \xi {\xi}^{\prime} \Pi - 2 r \xi  \Delta\Pi \right\}dr }
{ \int_{0}^{R} \rho \xi^{2} r^{2} dr} ~.
\ee

We will now compute the  frequency of oscillation 
for some anisotropic spheres. We will 
only consider models with $\gamma = const$. First, we
note that for stars with isotropic pressure and constant density, 
under a self-similar deformation $\xi={\rm const}\times r$,
\be
\omega_{0}^{2} =  4 \pi \rho_{0}(\gamma - \frac{4}{3})~.  
\ee
This result says that isotropic stars with $\gamma = 4/3$ are marginally 
stable. If  $ \gamma $   is less than $4/3$ then dynamical instability
will occur, while  if $\gamma $ is greater then $4/3$ the star is stable 
relative to the deformation $ \xi = r$.   
  
Computing $\omega$ from eq. (\ref{omega}) for the anisotropic model 
Case I, with the deformation $ \xi~ = ~r$,
we find

\be
\omega^{2} = 6 \rho_{0}
\left[(\frac{2 \pi}{3} - C)(\gamma - \frac{4}{3}) \right]
\ee

Thus, not surprisingly for this model, anisotropy results in a scaling of the 
frequency of oscillation. We had already seen earlier that the effect of 
the anisotropy was equivalent to a scaling of the density,
 and since the frequency of oscillation is proportional to the density,
this result is expected. Thus, for this case, positive anisotropy
may slow down the growth of instabilities, but will not reverse their
trend (recall that we must have $C\leq 2\pi/3$.)

We next consider the model Case II. For this model,

\be
\omega^{2} = 4 \pi \rho_{0}
 \left[ \gamma - \frac{4}{3} + \frac{16}{21}C \rho_{0} R^{2}
\right]
\ee
   
\noindent The fundamental frequency occurs for $\omega^{2} =0 $, and
this  corresponds here to 

\be
\gamma_c =  \frac{4}{3} - \frac{16}{21} C\rho_{0} R^{2}~.
\ee

\noindent  We 
observe that, since $ \rho_{0} R^{2}$ is always a positive quantity,
depending on the sign of  $C$  the value of $\gamma$ for which
 $\omega^{2}~ =~0$ can be less than or greater than  $4/3$, indicating 
that positive (negative) values of anisotropy can increase (decrease) 
the stability of the star.

\begin{figure}
\label{newfig}
\hspace{1.5in}
	\psfig{figure=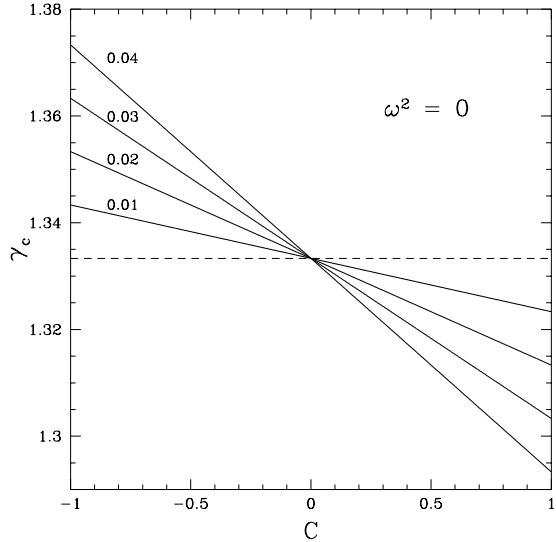,width=3.0in,height=3.0in}
	\caption[$ \gamma_c $ vs $ C $ for  $p_{t} - p_{r} = C \rho_0 p_{r} r^{2}$]
{Adiabatic index $\gamma_c$, for which $\omega^2=0$, as a function of $C$ for the ansatz 
 $p_{t} - p_{r} = C \rho_0 p_{r} r^{2}$ (Case II) parameterized by values of 
$\rho_{0}^{2}R$.}
\end{figure}

 We plot $ \gamma_c$  as a function of $C$ for $\omega^{2} = 0$ in
figure 3 for the ansatz  $p_{t} - p_{r} = C \rho p_{r} r^{2}$. 

This concludes our stability study of Newtonian anisotropic spheres. We have seen
that the presence  of anisotropic pressure in  a self-gravitating 
system can have dramatic effects on the dynamics and stability of the system.
In particular, there are some novel features that are present only if 
the pressure is anisotropic, e.g., infinite core pressure, zero radial pressure
and stable objects with $\gamma < 4/3$. We will now proceed to study the perturbation  
problem  for  relativistic  anisotropic compact spheres.

\section{Stability of General Relativistic  Anisotropic Spheres: General Formalism}

In this section we will study perturbations  of exact solutions of the
general relativistic field
equations for anisotropic spheres. In particular, we
will be concerned with perturbations that preserve spherical symmetry. Under
these perturbations, radial motions will ensue.
We will develop an analytical approach generalizing
work by Chandrasekhar for isotropic spheres \cite{CHANDRA}.

In this section $  \rho_{o}, \nu_{o}, \lambda_{o},
 {p_{t}}_{o}$ and ${p_{r}}_{o}$
are values of the dynamical
variables that satisfy the equations  for static equilibrium. The
perturbed variables will be written as  $ \rho, \nu , \lambda,
p_{t} $ and $p_{r}$, respectively.

\subsection{The Perturbed Energy-Momentum Tensor}

The energy-momentum tensor for a spherically symmetric spacetime is \cite{WEINBERG}

\be
\label{tmunu}
T_{\mu}^{\nu} = (\rho + p_{r})u_{\mu}u^{\nu} - g_{\mu}^{\nu}p_{r} 
- l_{\mu}l^{\nu}(p_{t} - p_{r})  - k_{\mu}k^{\nu}(p_{t} - p_{r})~.  
\ee
\noindent Here, 
\be
u^{\mu}  = \frac{dx^{\mu}}{ds}~,
\ee
\noindent and
\be
l_{\mu} =   {\delta}_{\mu}^{\theta},
~~~~ l^{\nu} = {\delta}_{\theta}^{\nu},~~~~ k_{\mu} = {\delta}_{\mu}^{\phi},
~~~~k^{\nu} = {\delta}_{\nu}^{\phi}.
\ee
\noindent Since we are considering only radial motions, we will take
\be
\label{ueq} 
u^{t} = e^{-\frac{{{\nu}}_{}}{2}},~~~~~~~~~~   
 u_{t} = e^{\frac{{{\nu}}_{}}{2}},
\ee
\noindent and
\be
\label{ueq2}
u^{r} = v e^{-\frac{{{\nu}}_{o}}{2}},~~~~~~~~~~
u_{r} = v e^{ \lambda_{o} -\frac{{{\nu}}_{o}}{2}},
\ee
\noindent with
\be
v = \frac{dr}{dt}.
\ee
It should be quite clear when a subscript refers to a time-like coordinate,
such as $u^t$, or tangential pressure, such as $p_t$. We reserve the index $o$ to
unperturbed metric and physical quantities.

Writing $ \rho ~ = ~ {\rho}_{o} ~ + ~\delta \rho,
~ p_{r} ~ = ~ {p_{r}}_{o} ~+ ~\delta p_{r},
~ p_{t} ~ = ~ {p_{t}}_{o} ~+ ~\delta p_{t},
~\lambda ~ = ~ {\lambda}_{o} ~ + ~\delta \lambda,
$ and $~\nu ~ = ~ {\nu}_{o} ~ + ~\delta \nu $
we find that, to first order in $v$ 

\be
T_{t}^{t} = \rho, ~~ T_{r}^{r} = -p_{r}, ~~ T_{\theta}^{\theta} ~=~ 
T_{\phi}^{\phi} 
= -p_{t} ~~ 
\ee
\noindent  and
\be
T_{t}^{r} = ( \rho_{o} + {p_{r}}_{o})v, ~~~~~~~~
T_{r}^{t} = ( \rho_{o} + {p_{r}}_{o})v e^{ (\lambda_{o} - \nu_{o})}. 
\ee

\subsection{ Perturbations of the Dynamical Variables}
The set of equations governing radial motions can be written as \cite{CHANDRA}:

\be
\label{iro1}
{\left( r e^{-\lambda} \right)}^{\prime} = 1 - 8 \pi \rho r^{2}~,
\ee
\be
\label{iro2}
{\nu}^{\prime} = r(e^{\lambda} - 1) + 8 \pi p_{r} r e^{\lambda}~,
\ee
\be
\label{iro3}
\frac{{\dot{\lambda}}}{r} e^{-\lambda} = 8 \pi T_{t}^{r}~,
\ee
\be
\label{iro4}
{\dot{T}}_{r}^{t} + \frac{1}{2} T_{r}^{t} \left( 
\dot{\lambda} + \dot{\nu} \right)  =  -  {p_{r}}^{\prime}   - 
\frac{1}{2} \left(\rho + p_{r} \right) {\nu}^{\prime}  + \frac{2}{r} \Pi
\ee
\noindent with
\be
\Pi \equiv\left(p_{t} - p_{r} \right).
\ee

The zeroth order (or static equilibrium) equations are:
\be
\label{iro5}
{\left( r e^{-{\lambda}_{o}} \right)}^{\prime} = 1 - 8 \pi {\rho}_{o} r^{2}~,
\ee

\be
\label{iro6}
{{\nu}_{o}}^{\prime} = r(e^{{\lambda}_{o}} - 1) + 
8 \pi {p_{r}}_{o} r e^{{\lambda}_{o}}~,
\ee
\noindent and
\be
\label{iro7}
{p_{r}}_{o}^{\prime}  = -({\rho}_{o} + {p_{r}}_{o})
\frac{{{\nu}}_{o}^{\prime}}{2} + \frac{2}{r} \Pi_{o}~.
\ee
\noindent We also have the identity
\be
\label{iro13}
\frac{e^{-\lambda_{o}}}{r} \left( {\lambda}_{o}^{\prime} + 
{\nu}_{o}^{\prime} \right)
= 8 \pi \left( {p_{r}}_{o} + {\rho}_{o} \right).
\ee

We now linearize eqs. (\ref{iro1})- (\ref{iro4}), taking into consideration
eqs. (\ref{iro5}) - (\ref{iro7}). Since we consider  all perturbations
to be of order $v$, we find that to first order in $v$,
eqs.  (\ref{iro1})- (\ref{iro4}) imply

\be
\label{iro8}
{\left( r e^{-{\lambda}_{o}} \delta \lambda \right)}^{\prime} =   8 \pi
 r^{2} \delta {\rho} ~,
\ee
\be
\label{iro9}
\frac{e^{{-\lambda}_{o}}}{r} \left(  { \delta{\nu}}^{\prime}
 - {{\nu}_{o}}^{\prime} 
\delta \lambda \right) = \frac{e^{{\lambda}_{o}}}{r^{2}} \delta \lambda + 
8 \pi \delta p_{r}~,
\ee
\be
\label{iro10}
\delta {\dot{\lambda}} \frac{e^{-\lambda}_{o}}{r} = -8 \pi \left(
\rho_{o}  +{ p_{r}}_{o} \right) v, 
\ee
\be
\label{iro11}
e^{{\lambda}_{o} - {\nu}_{o}} \left( {p_{r}}_{o} + {\rho}_{o} \right)
\dot{v}   +  {(\delta p_{r})}^{\prime}  + \frac{1}{2} \left(
{p_{r}}_{o} + {\rho}_{o} \right) \left( \delta \nu \right)^{\prime}   - 
\frac{1}{2} \left(\delta\rho + \delta p_{r} \right) {{\nu}}_{o}^{\prime} \\  
+ \frac{2}{r} \delta \Pi = 0.
\ee
  
We now introduce a ``Lagrangian displacement'' $\xi$ defined by
\be
v = \frac{\partial \xi}{\partial t}.
\ee
\noindent Integrating  eq. (\ref{iro10}),  we find that
\be
\label{iro14}
\delta {\lambda} \frac{e^{{-\lambda}_{o}}}{r} = -8 \pi \left(
\rho_{o}  +{ p_{r}}_{o} \right) \xi. 
\ee
\noindent Taking into consideration eq. (\ref{iro13}),
 the above equation becomes
\be
\label{iro15}
\delta \lambda = - \xi \left( \lambda_{o}^{\prime}  + \nu_{o}^{\prime} \right).
\ee
\noindent We can now combine eqs. (\ref{iro8}) and (\ref{iro15}) to get
\be
\delta \rho = - \frac{1}{r^{2}} \left[ r^{2} \left( {\rho}_{o} + {p_{r}}_{o}
\right) \xi \right]^{\prime}~.
\ee
\noindent Substituting for the expression for ${p_{r}}_{o}^{\prime}$ 
from eq. (\ref{iro7}) into the above equation, we find that
\be
\label{iro16}
\delta \rho = - \xi {{\rho}_{o}}^{\prime} - \left( \rho_{o} + {p_{r}}_{o}
\right) \frac{1}{r^{2}} e^{ {{\nu}_{o}}/2} \left( r^{2} e^{- {{\nu}_{o}}/2} \xi
\right)  - \frac{2 \xi}{r} \Pi_{o}~.
\ee

We now consider eq. (\ref{iro9}). Using eqs. (\ref{iro7}) and
(\ref{iro13}) we find

\be
\label{iro17}
\left( {p_{r}}_{o} + \rho_{o} \right) (\delta \nu)^{\prime} =
\left[ \delta p_{r}  - \left( {p_{r}}_{o} + \rho_{o} \right) 
\left( {\nu}_{o}^{\prime} + \frac{1}{r} \right) \xi \right] 
\left( \lambda_{o}  + \nu_{o} \right) ~.
\ee

We note that eqs. (\ref{iro15}), (\ref{iro16}) and (\ref{iro17}) allows us
to express  $ \delta \lambda$, $\delta \rho$,  and $\delta \nu$ in terms
of $\delta {p_{r}}$, $v$, and the unperturbed  variables.
We need to 
impose an extra 
condition on the system in order to obtain an expression for $\delta p_{r}$.
The condition we shall impose is the conservation of baryon number. Further,
we note that $\delta \Pi$ can always be expressed in terms of the unperturbed
variables once  $\delta p_{r}$ is given in terms of these variables. 

Chandrasekhar \cite{CHANDRA} derived an expression for $\delta p_{r}$ from
the law of conservation of baryon number in general relativity. Since we
are not making any new assumptions with respect to Chandrasekhar (except,
of course, we are considering anisotropic pressure),  we will 
only  outline the basic steps of his derivation here.

The law of conservation of baryon number density in general relativity can 
be written as
\be
\label{bar1} 
(n u^{\alpha})_{; \alpha} = 0~,
\ee

\noindent where $n$ is the number density for baryons and $u^{\alpha}$ is the 
four-velocity of the fluid. Taking 
\be
n = n_{o}(r) + \delta n(r,t)~,
\ee
\noindent  and recalling that, to first order in $v$, 
 $u^{\alpha}$ is given  by eqs. (\ref{ueq}) and (\ref{ueq2}),
  eq. (\ref{bar1}) becomes 
\be
\label{bar2}
e^{-\nu_{o}/2}
 \dot{\delta n} + 
\frac{1}{r^{2}} \left(n_{o} r^{2} v e^{-\nu_{o}/2} \right) +
\frac{1}{2} n_{o} e^{-\nu_{o}/2} \dot{\delta \lambda}   +
\frac{1}{2}e^{-\nu_{o}/2} v \left(\lambda_{o} + \nu_{o} \right)^{\prime} = 0.
\ee
\noindent Since $v = \dot{\xi}$, eq. (\ref{bar2}) integrates to give
\be
\label{bar3}
{\delta n} + 
\frac{e^{\nu_{o}/2}}{r^{2}} \left(n_{o} r^{2} 
v e^{-\nu_{o}/2} \right)^{\prime} +
\frac{1}{2}n_{o} \left[\delta \lambda   +
\xi \left(\lambda_{o} + \nu_{o} \right)^{\prime} \right] = 0.
\ee

\noindent The last term on the left-hand side of eq. (\ref{bar3}) vanishes
on account of eq. (\ref{iro15}), and we obtain
\be
\label{deltan}
{\delta n} =
- \frac{e^{\nu_{o}/2}}{r^{2}} \left(n_{o} r^{2} v 
e^{-\nu_{o}/2} \right)^{\prime}~. 
\ee
The first law of thermodynamics in general relativity is obtained by
combining
\be
u_{\nu} T^{\mu \nu}_{~~~;\mu} = 0,
\ee
\noindent with the law of conservation of baryon number given by
 eq. (\ref{bar1}).
Using  the expressions for $T^{\mu \nu}$  and $u_{\mu}$ from  
eqs. (\ref{tmunu}), (\ref{ueq}) and (\ref{ueq2}),
we find that,
\be
p_{r}d\left(\frac{1}{n}\right) + d\left( \frac{\rho}{n} \right) 
+ \frac{2}{r}(p_{t} - p_{r}) \frac{v}{n} dr = 0.
\ee
\noindent Thus, in general, the equation of state is given by
\be
n \equiv   n(\rho, p_{r}, \Pi) ~.
\ee
\noindent However, since we are considering systems where the  tangential 
pressure is given in terms of the
radial pressure and the density (recall that in generating exact solutions 
in the previous chapter we assumed various ansatze for $p_{t} - p_{r}$),
 we will take
\be
\label{stt}
n \equiv n(\rho, p_{r}) ~.
\ee
For this $n$, we have
\be
\delta n = \frac{\partial n_{o}}{\partial \rho} \delta \rho  +
 \frac{\partial n_{o}}{\partial p_{r}} \delta p_{r}  + 
 \frac{\partial n_{o}}{\partial r} dr.  
\ee
Substituting for $\delta n $ from eq. (\ref{deltan}) and $\delta \rho$ from
eq. (\ref{iro16}), we find that 
\bea
-\xi \frac{dn_{0}}{dr} - n_{o} \frac{e^{\nu_{o}}}{r^{2}} 
( r^{2} e^{\nu_{o}/2} \xi)^{\prime} &=&  
-\left( p_{{r}_{o}} + \rho_{o} \right) \frac{e^{\nu_{o}/2}}{r^{2}}
( r^{2} e^{\nu_{o}/2} \xi)^{\prime} \frac{\partial n_{0}}{\partial \rho} \\ \nonumber
&-&  \xi \frac{d\rho}{dr}\frac{\partial n_{0}}{\partial \rho} 
-\frac{2 \xi}{r} \Pi_{o}  \frac{\partial n_{0}}{\partial \rho} 
+ \frac{\partial n_{o}}{\partial p_{r}} \delta p_{r}  + 
 \frac{\partial n_{o}}{\partial r} dr.  
\eea
Dividing through out by $(\partial n_{o}/\partial p_{r})$ gives, to first order,
\bea
-\xi \frac{dp_{r_{o}}}{dr} - \frac{1}{\frac{\partial n_{o}}{\partial p_{r}}}
n_{o} \frac{e^{\nu_{o}}}{r^{2}} 
( r^{2} e^{\nu_{o}/2} \xi)^{\prime}
 &=&  
-\left( p_{{r}_{o}} + \rho_{o} \right) \frac{e^{\nu_{o}/2}}{r^{2}}
( r^{2} e^{\nu_{o}/2} \xi)^{\prime} \frac{\frac{\partial n_{o}}{\partial \rho}}{\frac{\partial n_{o}}{\partial p_{r}}}  \\ \nonumber
&-& \xi \frac{d\rho}{dr}
\frac{\frac{\partial n_{o}}{\partial \rho}}{\frac{\partial n_{o}}{\partial p_{r}}} 
 -\frac{2 \xi}{r} \Pi_{o} \frac{\frac{\partial n_{0}}{\partial \rho}}{
\frac{\partial n_{0}}{\partial p_{r}}} 
+  \delta p_{r}  + \frac{\frac{d n_{o}}{d r}}{
\frac{\partial n_{0}}{\partial p_{r}}}  dr.  
\eea
Solving for $\delta p_{r}$ we find
\be
\delta p_{r} =  - {p_{r}}^{\prime}_{o} \xi - 
 \frac{1}{{{p_{r}}_{o}}\frac{\partial n_{o}}{\partial p_{r}}} {p_{r}}_{o}
\left[ n_{o} - \frac{\partial n_{o}}{\partial \rho} (\rho_{o} + {p_{r}}_{o})
\right]  \frac{e^{\nu_{o}/2}}{r}
( r^{2} e^{-\nu_{o}/2} \xi)^{\prime}  + \frac{2 \xi}{r} \Pi_{o} \frac{\partial
{p_{r}}_{o}}{\partial \rho_{o}}.
\ee 
We can rewrite this as 
\be
\delta p_{r} = - {p_{r_{o}}}^{\prime} - \gamma {p_{r}}_{o} 
\frac{e^{\nu_{o}/2}}{r^{2}} \left( r^{2} e^{ {\nu}_{o}/2} \xi \right)^{\prime}
+ \frac{2 \xi}{r} \Pi_{o} \frac{\partial p_{r o}}{\partial \rho_{o}}~,
\ee
\noindent with $\gamma$ being the adiabatic exponent defined as
\be
\gamma \equiv \frac{1}{p_{r} (\partial n/\partial p_{r})} \left[ n - (\rho
+ p_{r})\frac{\partial n}{\partial p_{r}} \right]~.
\ee

\subsection{The Pulsation Equation}

We now assume that all perturbations have a time dependence of the form
$e^{i \omega t}$. Further, considering $\delta \lambda$, $ \delta \nu $,
$\delta \rho$, $\delta p_{r}$ and $\delta \Pi$ to now represent the amplitude
of the various perturbations with the same  time dependence we find, 
from eq. (\ref{iro11}), that
\bea
\omega^{2} \left( \rho_{o} + {p_{r}}_{o} \right) \xi e^{ \lambda_{o} - \nu_{o}}
 = \left(\delta p_{r} \right)^{\prime} + \delta p_{r} [ \frac{1}{2} 
\lambda_{o}^{\prime} + \nu_{o}^{\prime} ]  + \frac{1}{2} 
\delta \rho \nu_{o}^{\prime} \\ \nonumber
 - \frac{1}{2} \left(\rho_{o} + {p_{r}}_{o} \right) ( \nu_{o}^{\prime}
 + \frac{1}{r} ) \xi \left( \lambda_{o}^{\prime} 
+ \nu_{o}^{\prime} \right)
 - \frac{2}{r} \delta \Pi ~.
\eea

Substituting  the expressions for the  various amplitudes  in the above
equation we find
\bea
\label{iro23}
\lefteqn{\omega^{2} \left( \rho_{o} + {p_{r}}_{o} \right) \xi e^{ \lambda_{o} - \nu_{o}}
 =  -({p_{r}}_{o}^{\prime} \xi)^{\prime} 
- \frac{1}{2} \left(\rho_{o} + {p_{r}}_{o} \right) \left( \nu_{o}^{\prime}
 + \frac{1}{r} \right) \xi \left( \lambda_{o}^{\prime} 
+ \nu_{o}^{\prime} \right) } \\ \nonumber
& & - e^{-(\lambda_{o} + 2 \nu_{o})/2} 
\left[ e^{(\lambda_{o} + 2 \nu_{o})/2} \gamma {p_{r}}_{o} 
\frac{e^{\nu_{o}/2}}{r^{2}}(r^{2}e^{-\nu_{o}/2} \xi)^{\prime} \right]^{\prime} 
+ [\frac{1}{2} \lambda_{o}^{\prime}  
+ \nu_{o}^{\prime}]{{p_{r}}_{o}}^{\prime} \xi \\ \nonumber
& & - e^{-(\lambda_{o} + 2 \nu_{o})/2}\left[ e^{(\lambda_{o} + 2 \nu_{o})/2} 
\frac{2}{r} \xi \Pi_{o} \frac{\partial p_{r}}{\partial \rho} \right]^{\prime} 
- \frac{2}{r} \delta \Pi ~.
\eea

\noindent From eq. (\ref{iro7}) it follows that  
\be
\label{iro21}
{{p_{r}}_{o}}^{\prime \prime}  = -({\rho}_{o}^{\prime} + {p_{r}}_{o}^{\prime})
\frac{{{\nu}}_{o}^{\prime}}{2} + 
({\rho}_{o} + {p_{r}}_{o})\frac{{{\nu}}_{o}^{\prime \prime}}{2} + 
(\frac{2}{r} \Pi_{o})^{\prime}~,
\ee
\noindent and
\be
\label{iro20}
{\nu}_{o}^{\prime}  = \frac{-2 r{p_{r}}_{o}^{\prime} + 4 \Pi_{o}}
 {r({\rho}_{o} + {p_{r}}_{o})} ~.
\ee
\noindent Also, we have 
\be
\label{iro22}
\nu_{o}^{\prime \prime} - \nu_{o}^{\prime} \lambda_{o}^{\prime} - \frac{1}{r}
\lambda_{o}^{\prime} = 16 \pi (\Pi_{o} + {p_{r}}_{o})e^{\lambda_{o}} 
- \frac{1}{2} {{\nu_{o}}^{\prime}}^{2}  - \frac{1}{r} \nu_{o}^{\prime}~.
\ee

\noindent Using eqs.  (\ref{iro21}), (\ref{iro20}) and (\ref{iro22}) in 
(\ref{iro23}) we arrive at the pulsation equation i.e., 
the equation that governs radial oscillations:
\newpage

\bea
\label{iro24}
\lefteqn{\omega^{2}\left( \rho_{o} + {p_{r}}_{o} \right) \xi 
e^{ \lambda_{o} - \nu_{o}}
 = \frac{4}{r}{p_{r}}_{o}^{\prime} \xi
- e^{-(\lambda_{o} + 2 \nu_{o})/2} 
\left[ e^{(\lambda_{o} + 3 \nu_{o})/2} \gamma \frac{{p_{r}}_{o}}{r^{2}} 
(r^{2}e^{-\nu_{o}/2} \xi)^{\prime} \right]^{\prime} } +  \\  \nonumber
& & 8 \pi e^{\lambda_{o}}(\Pi_{o} 
+ {p_{r}}_{o})( \rho_{o} + {p_{r}}_{o}) \xi  
- \frac{1}{ (\rho_{o} + {p_{r}}_{o})} ({p_{r}}_{o}^{\prime})^{2} \xi 
 +  \frac{4 {{p_{r}}_{o}}^{\prime} \Pi_{o} \xi}{r( \rho_{o} + {p_{r}}_{o})}
-\frac{4 \Pi_{o}^{2}  \xi}{r^{2}( \rho_{o} + {p_{r}}_{o})} \\ \nonumber
& & - e^{-(\lambda_{o} + 2 \nu_{o})/2} 
\left[ e^{(\lambda_{o} + 2 \nu_{o})/2}\frac{2}{r} \xi \Pi_{o} 
\left(\frac{\partial p_{r}}{\partial \rho} + 1\right)  \right]^{\prime}   
 - \frac{8}{r^{2}}\Pi_{o}\xi 
 - \frac{2}{r} \delta \Pi ~.
\eea
\noindent The boundary conditions imposed on this equation are
\be
\label{iro25}
\xi = 0 {\rm ~at ~} r = 0 {\rm ~~~and~~~} \delta p_{r} = 0 {\rm ~at~} r = R.
\ee
 
The pulsation eq. (\ref{iro24}), together with the boundary conditions
eq. (\ref{iro25}), reduce to an eigenvalue problem for the 
frequency $\omega$ and
amplitude $\xi$. This is equivalent to the Sturm-Liouville problem
we encountered while studying  Newtonian gravity in chapter 2. Multiplying
eq. (\ref{iro24}) by $r^{2} \xi e^{(\lambda + \nu)/2}$ and integrating
over the entire range of $r$ we find that 
\begin{eqnarray}
\label{eqvalue}
\lefteqn{\omega^{2} \int_{0}^{R}   e^{ (3\lambda - \nu)/2} 
\left( \rho + p_{r} \right) r^{2} {\xi}^{2} dr = 
4 \int_{0}^{R}  e^{ (\lambda + \nu)/2} {p_{r}}^{\prime} r {\xi}^{2}
dr  } \\ \nonumber
& & +  \int_{0}^{R} e^{(\lambda + 3\nu)/2}\gamma \frac{{p_{r}}}{r^{2}} 
 \left[\left(r^{2}e^{-\nu/2} \xi \right)^{\prime}\right]^{2} 
- \int_{0}^{R}  e^{ (\lambda + \nu)/2}  \frac{r^{2} \xi^{2}}
{ (\rho + {p_{r}})} (p_{r}^{\prime})^{2} dr \\ \nonumber
& & 8 \pi \int_{0}^{R} e^{(3\lambda + \nu)/2}(\Pi 
+ {p_{r}})( \rho + {p_{r}}) r^{2} \xi^{2} dr    
 + 4 \int_{0}^{R} e^{ (\lambda + \nu)/2} 
\frac{p_{r}^{\prime} \Pi}{( \rho + p_{r})} r \xi^{2} dr \\ \nonumber
& & - 4 \int_{0}^{R} e^{ (\lambda + \nu)/2} 
\frac{ \Pi^{2}}{( \rho + p_{r})} \xi^{2} dr 
- 8 \int_{0}^{R} e^{ (\lambda + \nu)/2}  \Pi \xi^{2} dr 
- 2 \int_{0}^{R} e^{ (\lambda + \nu)/2}  \delta\Pi r \xi^{2} dr  \\ \nonumber
& & - \int_{0}^{R} e^{- \nu/2} 
\left[ e^{(\lambda + 2 \nu)/2}\frac{2}{r} \xi \Pi
\left(\frac{\partial p_{r}}{\partial \rho} + 1\right)  \right]^{\prime}  r^{2}
\xi dr ~,
\end{eqnarray}  
\noindent where we have dropped the subscripts as no longer necessary. The
orthogonality condition is now 
\be
 \int_{0}^{R}   e^{ (3\lambda - \nu)/2} 
\left( \rho + p_{r} \right) r^{2} {\xi}^{i} \xi^{j} dr = 0 
~~~~~~~~~(i \neq j)~,
\ee
\noindent where $ \xi^{i}$ and $\xi^{j}$ are the proper solutions belonging
to different eigenvalues  $\omega^{2}$.

\section{Stability of Anisotropic Spheres: $\rho \,=\, const.$ }
Chandrasekhar studied 
 the dynamical stability of isotropic spheres with
constant density $\rho$ and constant adiabatic index $\gamma$ using the
above formalism. The full
solution of Einstein's field equations for constant-density isotropic spheres 
is well
known \cite{WEINBERG}. Thus, writing 
\be
y = 1  - \frac{r^{2}}{\alpha^{2}}  \equiv  1  - \eta^{2}
~~~~~~{\rm and}~~~~~~
  y_{1} = 1 - \frac{R^{2}}{\alpha^{2}} \equiv 1 - \eta_{1}^{2}
\ee
\noindent  with 
\be
 \alpha^{2} = \frac{3}{8 \pi \rho},
\ee
\noindent the complete interior static isotropic solution 
for $ \rho = const$ is 
\be
\label{ior31}
p = \rho \left[ \frac{y - y_{1}}{3y_{1} - y} \right],
 ~~~~ e^{\lambda} = \frac{1}{y^{2}} ~~~~
{\rm and}~~~~ e^{\nu} = \frac{1}{4}(3y_{1} - y)^{2}~.
\ee
\noindent Here, we will apply the formalism just developed to the
Bowers-Liang solution for anisotropic spheres \cite{BOWERS},
\be
\label{ior32}
p_{r} = \rho \left[\frac{y^{2Q} - y_{1}^{2Q}}{3y_{1}^{2Q} - y^{2Q}} \right],
 ~~~ e^{\lambda} = \frac{1}{y^{2}} ~~~
{\rm and}~~~ e^{\nu} = \frac{1}{4}(3y_{1}^{2Q} - y^{2Q})^{1/Q}
\ee
\noindent and
\be
\label{nuuu2}
p_{t} - p_{r} = C \frac{\rho^{2} r^{2}}{y^{2}}
 \frac{4y^{2Q} y_{1}^{2Q}}{(3y_{1}^{2Q} - y^{2Q})^{2}},
\ee
\noindent with
\be
Q = \frac{1}{2} - \frac{3C}{4 \pi} \equiv  \frac{1}{2} - \frac{k}{2}.
\ee
\noindent In his stability analysis for isotropic stars,
Chandrasekhkar used the following trial function
\be
\label{trial}
\xi = \eta e^{\nu/2} = \frac{1}{2} \eta (3y_{1} - y)~,
\ee
\noindent and found that, to first order in $2M/R$,
the frequency of oscillation is given by
\be
\omega^{2} = \frac{1}{2 \alpha^{2}}[(3\gamma - 4) - \frac{1}{14}(\frac{2M}{R})
(54 \gamma -53)]~.
\ee
\noindent The first term reproduces  the results   
for Newtonian isotropic spheres [ cf. eq. (43)] and  
the second term represents a correction due to general
relativity.

We now turn our attention to the anisotropic case. A trial function that
generalizes (\ref{trial}) to include the effects of anisotropy
is 
\be
\xi_{c} = \eta e^{Q\nu}~,
\ee

However, we found that for the anisotropic case the corresponding 
integrals
in the expression for $\omega^{2}$ cannot be computed analytically if this
substitution is made. In order
to compute the integrals analytically, we used trial functions of
the following  form
\be
\label{trial22}
\xi_{1} = \eta^{\frac{1}{2}} e^{Q\nu}~,
\ee
\noindent  and
\be
\label{trial23}
\xi_{2} = \eta^{\frac{3}{2}} e^{Q\nu}~.
\ee

For the trial function $\xi_{1}$, we found, after integrating all terms
in eq. (\ref{eqvalue}), [for small $2M/R$]
\be
\omega^{2} = \frac{25}{16 \alpha^{2}} \left[ \gamma - \frac{32}{25} - k(\gamma
 - \frac{52}{25}) \right]  - \frac{25}{12 \alpha^{2}
} \left[ \gamma - \frac{23}{25}
-k(\frac{15}{8} \gamma - \frac{209}{100}) \right] 
\left(\frac{2M}{R} \right).
\ee
\noindent Thus, for this model, stable oscillations will occur if 
\be
\gamma \geq \frac{32}{25} - \frac{4}{5}k + \left[ \frac{36}{75}
- k \right] \left(\frac{2M}{R}\right)~.
\ee
\noindent For the trial function $\xi_{2}$ integrating eq. 
(\ref{eqvalue}) gives,
\be
\omega^{2} = \frac{49}{32 \alpha^{2}} \left[ \gamma - \frac{64}{49} - k(\gamma
- \frac{141}{49}) \right]  - \frac{245}{128 \alpha^{2}
} \left[ \gamma - \frac{48}{49}
-k(\frac{931}{490}\gamma - \frac{586}{245}) \right] \left(\frac{2M}{R} \right)
\ee 
\noindent and the condition for stable oscillations becomes
\be
\gamma \ge  \frac{64}{49} - \frac{11}{7}k +
 \left[ \frac{20}{49} - \frac{327}{196}k \right] \left(\frac{2M}{R}\right)~.
\ee
An examination of the two expressions for $\gamma$ above shows that for
positive anisotropy ($k ~>~ 0$) $\gamma$ is smaller than the 
corresponding isotropic value, implying that positive $k$ leads 
to more stable configurations, while negative values of $k$ 
will have a destabilizing effect. Further, we note that since for small values of $k$
the various new analytical solutions we found previously for constant-density anisotropic spheres 
\cite{KRSNA} have a 
similar form to the Bowers-Liang solution, we expect the relationship found here between
the sign of $k$ and the stability of the sphere to hold
also for those new solutions.   

It is useful to compare the results 
Chandrasekhar found  using the trial function $\xi$ [eq. (104)]
(denoted by $\gamma_{Ch}$)  ,
with the
results obtained with  our trial functions $\xi_{1} $ and $\xi_{2}$ 
(denoted by $\gamma_{1}$ and $\gamma_{2}$) in the isotropic limit ($k=0$).   
For stable oscillations  we must have
\bea
\gamma_{Ch} &\ge& \frac{4}{3} + \frac{19}{42}(\frac{2M}{R}) = 1.333
+ 0.4523 (\frac{2M}{R}) ~,\\ \nonumber
\gamma_{1}  &\ge& \frac{32}{25} + \frac{36}{75}(\frac{2M}{R}) = 1.280  
+ 0.480(\frac{2M}{R}) ~,\\ \nonumber
\gamma_{2}  &\ge& \frac{64}{49} + \frac{20}{49}(\frac{2M}{R}) = 1.306
+ 0.4081(\frac{2M}{R}).
\eea
\noindent It is known that Chandrasekhar's trial function becomes 
exact in the limit $(2M/R) \rightarrow 0$. Comparing the results for the
three trial functions above, we see that in the exact limit our results
differ from the exact value by $\approx 5 \%$. This leads us to believe that 
our trial functions generate
results that are qualitatively correct.  

\section{Stability of Anisotropic Spheres: $\rho \propto  1/{r^{2}} $ }
In \cite{KRSNA} we found several exact solutions for 
 anisotropic stellar configurations with
the following
expression for the energy density
\be
\label{density}
\rho = \frac{1}{8 \pi} \left( \frac{a}{r^{2}} + 3b \right),
\ee
\noindent where both  $a$ and $b$ are constants. 
The choice of the values for 
$a$ and $b$ is dictated by the physical configuration under consideration.
For example, $a=3/7$ corresponds to the Misner-Zapolsky
solution for ultra high-density neutron star cores \cite{MISNER}.

If we model   the pressure anisotropy as
\be
\label{ansio}
p_{t}  - p_{r} = \frac{1}{8 \pi}\left( \frac{c}{r^{2}} + d \right)~,
\ee
then an exact solution  of the field equations with $b \; = \; d = \;0$
is 

\be
e^{-\lambda} = 1 - a \; = I_{0}
\ee
\be
e^{\frac{\mu}{2}}= A_{+} \left(\frac{r}{R} \right)^{1+q} + 
A_{-} \left(\frac{r}{R} \right)^{1-q}~,
\ee

\noindent with

\be
q \equiv  \frac{( 1 +c - 2a)^{\frac{1}{2}}}{(1 - a)^{\frac{1}{2}}}~,
\ee
and the constants $A_{+}$ and $A_{-}$ fixed by boundary conditions. 
For the case under consideration here ($ b = d=0$), the 
boundary conditions are
\be
e^{-\lambda(R)} = e^{\nu(R)}  =  I_{0}^{2}, ~~~{\rm and}
~e^{\nu (R)} \frac{d \nu}{dr}|_R = \frac{a}{R}~.
\ee
 Applying the boundary conditions we find 
\be
A_{+} = \frac{I_{0}}{2} + \frac{1 - 3I_{0}^{2}}{4qI_0} ~~~~~ {\rm and} 
~~~~~A_{-} = A_{+}(q \rightarrow - q).
\ee

\noindent The radial pressure for $q$ real, after substituting the expressions
for $A_{+}$ and $A_{-}$, is 
\be
8 \pi p_{r} = \frac{(3I_{0}^{2} - 1)^{2} - 4q^{2}I_{0}^{4}}{r^{2}}
\left[ \frac{R^{2q} - r^{2q}}{(3I_0^{2} - 1 + 2qI_0^{2})R^{2q}  
+ ( 1 -3I_0^{2} + 2qI_0^{2})r^{2q}} \right]~.
\ee

For this case we found that using  the following trial function 
\be
\xi  = r^{2} ( \rho + p_{r}) e^{\nu}
\ee
\noindent all the integrals were, after some
tedious work, exactly integrable. In table I,
we present results for the frequencies of radial oscillations 
 ${\omega}^{2}$ as a function of the anisotropy parameter, $c$, for 
given values of the density parameter. We also give, in table II, the values of
$\gamma_{c}$ above  which stable oscillations are possible. Here we
see that the effect of a positive anisotropy is to reduce the value
of $\gamma$, thus giving rise to a more stable configuration when compared
with the corresponding isotropic model. In particular, for the Misner-Zapolsky
solution ($a=3/7$), we find that a small positive pressure anisotropy in the equation
of state improves the neutron star's core stability.

\begin{table}
\begin{center}
\begin{tabular}{|l|l|r|}  \hline

a =  2/9  & $\omega^{2}R^{2}$ =   0.95($\gamma$ -1.79) 
+ (101.1 -52.6  $\gamma$ )c \\  \hline

a = 2/7  &  $ \omega^{2}R^{2}$ =   2.3($\gamma$ - 1.83) 
+ (122.3 - 59.3 $\gamma$)c \\ \hline

a = 3/7  &   $\omega^{2}R^{2} $=   0.57($\gamma$ -    1.93) +
(15.2 - 5.1$ \gamma$)c  \\ \hline

a = 3.4/7 & $\omega^{2}R^{2}$ = 0.4($\gamma$ - 2.6)
+(8.9 - 2.3 $\gamma$)c \\  \hline
a = 3.49/7  &   $\omega^{2}R^{2}$ =   0.36($\gamma$ - 2.76) 
+ (8.0 -1.97 $\gamma$)c  \\ \hline

\end{tabular}
\caption{ $\omega^{2}$ $vs.$   $c$  for given values of $a$.}

\end{center}
\end{table}

\begin{table}
\begin{center}
\begin{tabular}{|l|l|l|r|}  \hline

a =  2/9  & $c_{max}$ = 0.0016 &$\gamma_{c}$ = 1.79 -6.87 c \\  \hline

a =  2/7  & $c_{max}$ = 0.0028 &$\gamma_{c}$ = 1.83 - 13.39 c \\  \hline

a =  3/7  & $c_{max}$ = 0.083 &$\gamma_{c}$ = 1.93 - 5.55 c \\  \hline

a =  3.4/7  & $c_{max}$ = 0.11 &$\gamma_{c}$ =  2.6 - 2.84 c \\  \hline

a =  3.47/7  & $c_{max}$ = 0.12 &$\gamma_{c}$ = 2.75 - 7.29 c \\  \hline

\end{tabular}
\caption{ $\gamma_{c}$ $vs$   $c$  for given values of $a$.}

\end{center}
\end{table}

\section{Conclusion}
We have studied the stability of anisotropic compact spheres 
against radial perturbations
in  the framework of  Newtonian gravity and general relativity. In
both cases we have seen that the presence of anisotropic pressure
can have significant effects.   

We have found that there are Newtonian anisotropic spheres with constant
energy density whose core
pressure can go to infinity, without requiring the radius of the sphere to
be infinite. A result of this nature is not possible  for Newtonian isotropic
spheres with constant energy density. Furthermore, a stability analysis of 
some of these models shows that there can exist stable anisotropic spheres
with an  adiabatic exponent $\gamma
 < 4/3$. In the corresponding isotropic case, instability immediately
sets in if $ \gamma < 4/3$.
   
We have extended the formalism developed by Chandrasekhar to  
study the stability of general relativistic isotropic spheres  
against radial perturbations to anisotropic spheres. In particular, we 
have applied this formalism to study anisotropic spheres with constant
energy density and with energy densities with 
an $1/r^2$ profile, used to model ultra-dense neutron star interiors,
for example. We have found that in both cases there can exist stable relativistic 
anisotropic spheres with values of the adiabatic exponent 
that would necessarily imply instability in isotropic spheres. 
In particular, this is true whenever the tangential pressure
is larger than the radial pressure for all models we investigated.
These results may explain
the higher stability of certain neutron stars with anisotropic deviations
near their core, and other gravitationally-bound compact objects such as boson stars,
which are naturally anisotropic.
Work along these lines is currently in progress.

\vspace{1.cm}

\noindent
{\bf Acknowledgements}
MG was supported in part by NSF grant
PHY-0099543.

\end{document}